\documentclass[a4paper, 11pt]{article}
\usepackage[colon]{natbib}
\usepackage{tabularx}
\usepackage{epsfig}
\usepackage[titletoc, title]{appendix}
\usepackage{setspace}
\usepackage{lscape}
\usepackage{booktabs}
\usepackage{array}
\usepackage{tabularx}
\usepackage[english]{babel}
\usepackage{makeidx}
\usepackage{latexsym}
\usepackage{ifthen}
\usepackage{amssymb}
\usepackage{color,epsfig}
\usepackage{graphicx}
\usepackage{subfig}
\usepackage{mathrsfs}
\usepackage{wrapfig}
\usepackage{rotating}
\usepackage{natbib}
\usepackage{amsmath, amsthm}
\usepackage{nomencl}
\usepackage{float}
\usepackage{multirow}
\usepackage{multicol}
\usepackage{xcolor}
\usepackage{epic,eepic}
\usepackage{lscape}
\usepackage{tabularx}
\usepackage{pdflscape}
\usepackage{epstopdf}
\usepackage{calligra}
\usepackage{epsfig}
\usepackage[colon]{natbib}
\usepackage{tabularx}
\usepackage{amsmath}
\usepackage{amssymb}
\usepackage{pdflscape}
\usepackage{multirow}
\usepackage{multicol}
\usepackage[bottom]{footmisc}
\usepackage{array,arydshln}
\usepackage{appendix}
\usepackage{hyperref}

\newcolumntype{x}[1]{>{\centering\arraybackslash\hspace{0pt}}p{#1}}

\DeclareMathSizes {8}{8}{8}{8}
\makeindex           
\allowdisplaybreaks

\newcommand{\bs}{\boldsymbol}
\newcommand{\tbf}{\textbf}

\begin{document}

\title{Latent drop-out based transitions in linear quantile hidden Markov models for longitudinal responses with attrition
}

\title{Latent drop-out transitions in quantile regression}        

\author{Maria Francesca Marino\thanks{Dipartimento di Econimia, Universit\`a degli Studi di Perugia, Italy, {\tt mariafrancesca.marino@uniroma1.it}} \and Marco Alf\'o\thanks{Dipartimento di Scienze Statistiche, Sapienza Universit\`a di Roma, Italy, {\tt marco.alfo@uniroma1.it}}}


\date{}

\maketitle

\begin{abstract}
Longitudinal data are characterized by the dependence between observations coming from the same individual. In a regression perspective, such a dependence can be usefully ascribed to unobserved features (covariates) specific to each individual. On these grounds, random parameter models with time-constant or time-varying structure are well established in the generalized linear model context. 
In the quantile regression framework, specifications based on random parameters have only recently known a flowering interest. We start from the recent proposal by \cite{Farcomeni2012} on longitudinal quantile hidden Markov models, and extend it to handle potentially informative missing data mechanism. In particular, we focus on monotone missingness which may lead to selection bias and, therefore, to unreliable inferences on model parameters. We detail the proposed approach by re-analyzing a well known dataset on the dynamics of CD4 cell counts in HIV seroconverters and by means of a simulation study.

\end{abstract}

\section{Introduction}
\label{intro}
Quantile regression has become a standard tool to model the distribution of a continuous response variable as a function of a set of observed covariates. When the interest lies not only on the center of the response distribution and/or when the observed data may include some outliers, quantile regression represents an interesting alternative to standard mean regression. During the last few years, the basic homogeneous quantile regression model (\citealp{KoenkerBassett1978}) has been extended to deal with longitudinal responses. To handle the dependence between measurements taken over time on the same individual, unit-specific, time-constant, random parameters can be added to the model specification (see eg \citealp{GeraciBottai2007, GeraciBottai2013}). A potential alternative is to consider time-varying random parameters. In this perspective, by extending standard hidden Markov models \citep{Wiggins1973}, \cite{Farcomeni2012} proposes a linear quantile hidden Markov model with a random intercept that varies over time according to a first-order hidden Markov chain. For a general treatment of hidden Markov models (HMMs) for longitudinal data, see \cite{BartFarcoPennoni2013} and references therein.
\\
\indent
A common feature of longitudinal studies is that individuals may leave the study before its end. Thus, variable-length individual sequences represent a further challenge, since not all individuals have the same weight in building up the log-likelihood function. A major problem is the so-called informative missingness: once conditioning on the observed covariates and responses, the selection of units in the study may still depend on future, unmeasured, responses. When ignored, this missing data generating mechanism may severely bias parameter estimates and lead to misleading conclusions. 
Following the proposals by \cite{Roy2003} and \cite{Roy2008}, we consider a pattern mixture representation \citep{Little1993} and develop a linear quantile hidden Markov model with latent drop-out classes. The idea behind such model is that, after conditioning on the observed covariates, differences between sample units arise due to unobserved heterogeneity. Random parameters varying over time according to a hidden Markov chain capture differences related to the dynamics of omitted covariates. A further source of unobserved heterogeneity may be represented by sub-samples of individuals being characterized by a different propensity to drop-out from the study. These sub-populations are identified by adding in the model a latent multinomial variable, whose ordered categories directly influence the Markov transition matrix. 
\\ \indent
The paper is structured as follows: in section 2, the linear quantile hidden Markov model is briefly reviewed. In section 3, we extend this proposal in a pattern mixture  perspective, by considering latent drop-out classes to capture individual-specific propensities to leave the study. The modified EM algorithm for parameter estimation is discussed in section 4; the proposed method is applied in section 5 to a well-known benchmark multi-center longitudinal study on the time progression of CD4 cell numbers in HIV seroconverters. Section 6 discusses the results of a simulation study. Last section contains concluding remarks and outlines potential, future, research lines.

\section{Linear quantile hidden Markov models }
Let us suppose a longitudinal study collects repeated measures of a \textit{continuous} response variable $Y_{it}$ on a sample of $i = 1,\dots, n$ subjects at time occasions $t = 1, \dots, T$. To account for dependence between measurements on the same statistical unit, a standard approach is to specify a conditional model for the responses, which are assumed to be independent, conditional on a set of individual-specific latent variables. In the context of generalized linear models for longitudinal responses, such latent effects may be either time-constant, as in mixed models \citep{LairdWare1982}, or time-varying, as in hidden Markov models \citep{Wiggins1973}. For a combination of both, see \cite{Altman2007} and \cite{Maruotti2011}. While this class of models has quite a long history in the generalized linear model framework, only recently its scope has been broadened to quantile regression, see \cite{GeraciBottai2007} and \cite{GeraciBottai2013}. Models with time-varying parameters have been introduced by \cite{Farcomeni2012} to model the (conditional) quantiles of a longitudinal response. This proposal (in the following lqHMM) is based on the existence of two related processes: a latent process with a Markov structure and an observed measurement process, whose parameters are defined by the current state of the hidden Markov chain. Conditional on the state occupied at a given time occasion, the longitudinal observations from the same individual are assumed to be independent (local independence assumption).
\\ \indent
Let us consider a quantile $\tau \in (0, 1) $, and denote by $S_{it}(\tau)$ a quantile-specific, homogeneous, first order, hidden Markov chain. The chain takes values in the finite set $\mathcal S(\tau) = \{1,\dots, m(\tau)\}$; $\bs \delta(\tau)$ and $\tbf Q(\tau)$ are the initial probability vector and the transition probability matrix of the chain, respectively.  The lqHMM can be specified as follows:
\begin{eqnarray}\label{lqHMMeq}
& Y_{it}  \mid s_{it} \sim {\rm ALD}(\mu_{it}(s_{it}, \tau), \sigma(\tau), \tau) \nonumber \\
& \mu_{it}\left(s_{it}, \tau \right)  =  \alpha(s_{it}, \tau)  + \tbf x_{it} ^\prime \bs \beta(\tau)
\end{eqnarray}
where $\mu$, $\sigma$ and $\tau$ are the location, the dispersion and the skewness parameters for the asymmetric Laplace distribution. The location parameter is linear in the time-varying intercept, $\alpha(s_{it},\tau)$, and in the vector of fixed effects $ \bs \beta( \tau)$. The assumption that the response variable has an asymmetric Laplace distribution, see \cite{GeraciBottai2007}, is made to recast standard quantile loss optimization within a maximum likelihood perspective. Moving from the random intercept to the more general random coefficient framework, we may write
\begin{equation*}
\mu_{it} \left(s_{it}, \tau \right) = \tbf x_{it} ^\prime \bs \beta(\tau) + \tbf z_{it}^\prime \bs \alpha(s_{it}, \tau) 
\end{equation*}
where $\bs \beta(\tau)$ summarizes the fixed effect of observed covariates on the $\tau$-th (conditional) quantile of the response distribution, while $\bs \alpha(s_{it}, \tau) $ represents the individual-specific effect associated to a subset of $\tbf x_{it}$ for an individual in state $s_{it}$ at time occasion $t$. 
Based on the modelling assumptions, the individual contribution to the observed data likelihood can be written as follows:
\begin{equation}\label{loglikFarco}
f_{Y} (\tbf y_i) = 
\sum_{\tbf s_i} f_{Y \mid S} (\tbf y_{i} \mid \tbf s_i) f_{S} (\tbf s_i)
\end{equation}
Obviously, this framework leads to quite a general structure of association between longitudinal measurements. However, this model can not properly handle incomplete sequences due to an informative missing data process \citep{LittleRubin2002}. In the next section, we extend such a model specification to account for individual differences in the propensity to leave the study.

\section{Handling informative missingness}
Let us consider a measurement process affected by monotone missingness: for each unit $i=1,\dots, n$, the measurements are available at time points $t = 1,\dots, T_i$ only, with $T_{i} \le T$. Let us denote by $R_{it}$ the missing data indicator variable, which is equal to 1 if the $i$-th subject is not available at the $t$-th occasion. Since we consider monotone missingness, $R_{it}=1 \Rightarrow R_{it^{\prime}}=1$, $t^{\prime}\geq t=1,\dots,T$. When the drop-out is informative, the missing data process needs to be properly modelled, at the risk of obtaining unreliable parameter estimates. 
The drop-out is defined to be informative when, conditional on observed responses and covariates, the missing data process still depends on the current, unobserved, values and/or when parameter distinctiveness between the distribution of $Y$ and $R$ does not hold, see \cite{LittleRubin2002} for a general treatment.
\\
\indent
In these cases, a more general model should be defined.
Few attempts to handle informative missingness have been made in the quantile regression framework; \cite{Lipsitz1997} and \cite{YiHe2009} suggest a GEE approach \citep{Liang1986}, while \cite{FarcoViviani2014} consider a joint model (JM) representation, see \cite{Rizo2012}. While this latter approach is an elegant way to handle dependence between longitudinal responses and missingness, JMs require the distribution for the missing data process to be completely specified and this often represents a delicate matter.
Here, we focus on pattern mixture models (PMMs), see \cite{Little1993}. The rationale for pattern-mixture models is that each subject has its own propensity to drop-out from the study. Individuals with similar propensities share some common observed/unobserved features and the model for the longitudinal response is given by a mixture over these patterns. Pattern mixture models do not need the distribution of the missing data generating process to be specified, but are often overparameterized. This issue may be (at least potentially) solved by defining appropriate identifying restrictions. Latent drop-out (LDO) models \citep{Roy2003, Roy2008} represent a viable solution. In such a specification, a limited number of LDO classes is considered to avoid overparametrization; sample units belonging to the same LDO class share common unobserved characteristics that influence, either directly or indirectly, the response variable distribution. 
To explain our proposal, let $\bs \zeta_i(\tau) = (\zeta_{i1}(\tau), ..., \zeta_{iG}(\tau))$ be a (quantile-specific) multinomial random variable with component $\zeta_{ig}(\tau)=1$ if subject $i$ belongs to the $g$-th drop-out class and zero otherwise. These categories represent ordered propensities to drop-out; that is, we assume that, for $g > g^\prime$, the propensity to drop-out for units with $ \zeta_{ig}(\tau) =1$ is lower than the propensity of units with $ \zeta_{ig^\prime}(\tau) = 1$. For a generic quantile $\tau \in (0,1)$, this ordering is specified through the following model:
\begin{equation}
\Pr \left(\sum _{l =1}^g \zeta_{il}(\tau) = 1 \mid T_i \right) = \frac{\exp \{ \lambda _{0g}(\tau) + \lambda _{1}(\tau)\, T_i\} }{1+ \exp \{ \lambda _{0g}(\tau) + \lambda _{1}(\tau)\, T_i\}}.
\end{equation}
under the constraint $\lambda _{0g}(\tau) \le \lambda _{0g\prime}(\tau)$ if $g<g^\prime$. 
The probability of belonging to one of the first $g$ classes is, thus, modelled as a monotone function of the time to drop-out; the probability of a specific class is obtained as the difference between two adjacent cumulative logits \citep{Agresti2010}. We prefer the proportional-odds specification used in \cite{Roy2008} over the non-proportional-odds discussed by \cite{Roy2003} since the common slope  for the curves defining the cumulative probabilities and the above constraints imply that the distribution of ${\bs \zeta}_{i}(\tau)$ at different values of $T_{i}$ is stochastically ordered. We assume that the latent drop-out class variable summarizes all the dependence between the longitudinal response and the missing data mechanism; conditional on the drop-out class, the two processes are independent. 
As it is obvious, LDO classes may influence the response variable distribution in several ways: for example, they may produce class-specific changes to the fixed effect parameter vector, as in \cite{Marino2014}. Alternatively, they may produce changes in the locations of the hidden Markov chain, thus giving rise to a LDO-specific support for the time-varying random parameters. Here, we discuss a further alternative; we assume that LDO representation produces a change in the matrix of transition probabilities. That is, we consider a (quantile-specific) homogeneous, first order, hidden Markov chain, $S_{it}(\tau)$, taking values in the finite set $\mathcal S (\tau)= \{1,\dots, m(\tau)\}$. The corresponding initial probability vector is assumed to be constant among LDO classes and is denoted by $\bs \delta(\tau)$, while the transition probability matrix $\tbf Q(g;\tau)$ is specific to each LDO class, $g=1,\dots, G$. This approach shares some features with the proposal by \cite{MaruottiRocci2012}; here, latent class-specific transitions are considered in the framework of standard HMMs.
As it is clear, the proposed specification covers a range of situations which is more general than a simple change in the location parameters of the hidden Markov chain. By allowing $\tbf Q(\cdot)$ depend on $g$, we may define states that are ``visited'' only by individuals in a given LDO class, that is latent class-specific parameter values.
The proposed model is in line with \cite{Bartolucci2015} and \cite{Maruotti2015}, where standard HMMs are extended to deal with informative drop-outs. More in detail, \cite{Bartolucci2015} discuss a shared parameter model with time-constant and time-varying (discrete) random intercepts shared by the longitudinal and the missing data process. \cite{Maruotti2015} describes a pattern mixture approach with the Markov transition matrix being a function of the time to drop-out. When compared with the former, our proposal does not need the distribution of the missing data process to be specified, thus allowing to avoid unverifiable parametric assumptions. When compared with the latter, our approach seems to be more general and offer greater flexibility. 
\\
\indent
Let $\bs \Psi(\tau) = (\boldsymbol \theta(\tau),  \sigma(\tau), \bs \delta(\tau), \tbf Q(\tau), \bs \lambda(\tau))$, where $\bs \theta(\tau) = (\bs \beta(\tau), \bs \alpha_1(\tau), \dots, \\\bs\alpha_{m(\tau)}(\tau))$ denotes the vector of longitudinal model parameters, and let $\bs \Phi(\tau)$  be the vector of parameters indexing the distribution of the time to drop-out, $f_T(T_i\mid \bs \Phi; \tau)$. Based on the previous modelling assumptions, the observed individual likelihood for a generic sample unit is obtained by marginalizing the joint distribution of the observed and the latent variables over the hidden Markov chain and the latent drop-out class indicator. Suppressing the dependence on model parameters to simplify the notation, the following expression holds:
\begin{equation}\label{loglikMarino}
f_{YT} (\tbf y_i, T_i; \tau) = 
\sum_{\tbf s_i \, \bs \zeta_i} 
f_{Y \mid S \zeta} (\tbf y_{i} \mid \tbf s_i, \bs \zeta_i; \tau) 
f_{S} (\tbf s_i \mid \bs \zeta_i; \tau) f_{\zeta \mid T} (\bs \zeta_i \mid T_i; \tau) f_T(T_i; \tau).
\end{equation}
From the above equation, it is clear that the marginal distribution of the time to drop-out can be left unspecified and ignored when maximizing the likelihood with respect to $\bs \Psi(\tau)$; inference may be based on the conditional distribution $f_{Y \mid T} (\tbf y_i \mid T_i; \tau)$ only.

\section{Parameter estimation}
The general structure of the EM algorithm \citep{Dempster1977} we use for parameter estimation can be sketched as follows. To keep the notation simple, we will omit the dependence of model parameters on the specific quantile $\tau$ we consider.
Let $u_{it} (h) = I(S_{it} = h)$ be the variable indicating the $i$-th unit is in the $h$-th hidden state at occasion $t$ and let $u_{it}(h,k)$ be the indicator variable for the $i$-th unit moving from the $h$-th state at occasion $t-1$ to the $k$-th one at $t$. Last, let $\zeta_{ig}$ be the indicator variable for unit $i= 1, \dots,n$ in the $g$-th latent class.
For a given quantile $\tau$, the (conditional) log-likelihood for  complete data is
\begin{align}
\ell_{c}(\bs \Psi) &= 
\sum_{i=1}^n \left\{
\sum_{h = 1}^m u_{it}(h) \log \delta_h + 
\sum_{t = 1}^{T_i} \sum_{h = 1}^m \sum_{k = 1}^m \sum_{g = 1}^G u_{it}(k,h) \zeta_{ig} \log q_{kh}(g) + \right.  \nonumber\\
& \left.  
+ \sum _{g = 1}^G \zeta_{ig} \log \pi_g  -  T_i \log \sigma 
-\sum_{t = 1}^{T_i} \sum_{h = 1}^m 
u_{it} (h) \rho_{\tau} \left( 
\frac{ y_{it} - \mu_{it}(S_{it} = h) }{\sigma} 
\right)  
\right\}
\end{align}
The E-step of the algorithm requires the computation of the expected values for the indicator variables $u_{it}(h), u_{it}(h,k)$ and $\zeta_{ig}$, conditional on the observed data and the current parameter estimates.  As it is usual with hidden Markov models, such a computation is simplified by considering the forward and backward variables \citep{Baum1970}. In the present framework, for a generic individual in the $g$-th latent drop-out class, forward variables, $a_{it} (h, g)$, define the joint density of the longitudinal measures up to time $t$ and the $h$-th state at $t$:
\begin{align}
a_{it}(h, g) =
f \left[y_{i1:t}, S_{it} = h \mid \zeta_{ig} = 1\right ].
\end{align}
Following \cite{Baum1970}, these terms can be computed recursively
\begin{align}\label{fwdRecursion}
& a_{i1}(h,g)  = \delta_h f_{Y \mid S} \big[ y_{i1} \mid S_{i1} = h\big]   \\ \nonumber
&a_{it}(h,g)  =
\sum_{k=1}^m a_{it-1}(k, g) q_{kh}(g) f_{Y \mid S} \big[ y_{it} \mid S_{it} = h \big].
\end{align}
Similarly, the backward variables, $b_{it} (h, g)$, represent the probability of the longitudinal sequence from occasion $t+1$ to the last observation, conditional on being in the $g$-th LDO class and in the $h$-th state at time $t$:
\begin{align}
b_{it}(h,g) =
f \big[ y_{it+1:T_i} \mid S_{it} = h, \zeta_{ig} = 1\big].
\end{align}
As for the forward, also backward variables can be derived recursively:
\begin{align}\label{bwdRecursion}
& b_{iT_{i}}(h,g) = 1 \\ \nonumber
& b_{it-1}(h,g) = \sum _{k=1}^m
b_{it}(k,g) q_{hk}(g) f_{y \mid sb} \big[ y_{it} \mid S_{it} = h \big] ,
\end{align}
For a detailed description of the Baum-Welch algorithm, see the seminal paper by \cite{Baum1970} and the reference monograph by \cite{Zucchini2009}.
\\
\indent
Computation of the expected complete data log-likelihood, conditional on the observed data and the current parameter estimates, leads to
\begin{align} \label{auxFnc_NPML}
& Q(\bs \Psi \mid \hat{\bs \Psi})  =
\sum _{i = 1} ^ n \bigg\{
\sum_{h = 1}^m \hat u_{i1}(h) \log \delta_{h} +
\sum_{t=2}^{T_i} \sum_{h,k=1}^m \sum_{g = 1}^G \hat \zeta_{ig} \hat  u_{it}(k,h\mid g) \log q_{kh} (g)+
\nonumber \\[0.1cm]
& \quad +\sum_{g =1}^G \hat \zeta_{ig} \log \pi_g 
- T_i \log (\sigma)
-\sum _{t = 1}^{T_i}  \sum_{h=1}^m  
\sum_{g=1}^G \Big[
\hat u_{it}(h)
\rho_{\tau} \left( \frac{ y_{it} - \mu_{it}(S_{it} = h)
} {\sigma} \right)
 \bigg\},
\end{align}
where $\hat u_{it}(h)$ and $\hat \zeta_{ig}$ represent the posterior expectation of the indicator variables we have previously introduced.
Moreover, $\hat u_{it}(k,h\mid g)$ denotes the posterior probability for the $i$-th unit in state $k$ at occasion $t-1$ and moving to state $h$ at occasion $t$, given she/he belongs to the $g$-th LDO class.
These posterior probabilities can be easily obtained by exploiting the forward and backward variables (\ref{fwdRecursion}) and (\ref{bwdRecursion}) as:
\begin{align}
& \hat u_{it}(h) = \frac
{
\sum_g
a_{it}(h,g)
b_{it}(h,g)
\pi_g
}
{
\sum_{h} \sum_g
a_{it}(h,g)
b_{it}(h,g)
\pi_g
} 		\nonumber \\[0.2cm]
 & \hat u_{it}(k,h \mid g) =
\frac{
a_{it-1}(k,g)
q_{kh}(g) \,
f_{s \mid s}\left(y_{it} \mid S_{it} = h,\right)
b_{it}(h,g)
} {
\sum_{h} \sum_k
a_{it-1}(k,g)
q_{kh}(g) \,
f_{Y \mid S}\left(y_{it} \mid S_{it} = h,\right)
b_{it}(h,g)
}. 	 \nonumber \\[0.2cm]
& \hat \zeta_{ig} = \frac{\sum_h a_{iT_{i}}(h, g) \pi _g}{\sum_g  \sum_h a_{iT_{i}}(h, g) \pi _g} \nonumber
\end{align}
The M-step of the EM algorithm require the maximization of the $Q(\cdot)$ function with respect to model parameters. Closed form solutions are available for the parameters of the hidden Markov process:
\begin{align} 
\hat \delta _h = \frac{\sum _{i = 1}^n \hat u_{i1}(h)}{n},
\quad
\hat q_{kh}(g) = \frac {\sum _{i = 1}^n \sum_{t=1}^{T_i} \hat u_{it}(k,h \mid g)}{\sum _{i = 1}^n \sum_{t=1}^{T_i} \sum _{h = 1}^m \hat u_{it}(k,h \mid g)}
\end{align}
The scale parameter of the longitudinal response distribution is estimated as
\begin{equation}
\hat \sigma = \frac{1}{\sum_{i=1}^n T_i} \sum_{t=1}^{T_i} \sum_{h=1}^m  \hat{u}_{it}(h) \rho_{\tau} 
\left( 
y_{it} - \hat \mu_{it}(S_{it}=h)
\right).
\end{equation}
Parameters in the longitudinal and in the LDO class model, ($\bs \theta, \bs \lambda$), are estimated by finding the zeros of weighted score functions. For the longitudinal outcome, weights are given by the posterior probabilities of the hidden states, $\hat u_{it}(h)$; the following estimating equation holds
\begin{align}
\sum_{i=1}^n \sum _{t = 1}^{T_i}  \sum_{h=1}^m \hat u_{it}(h)
\frac{\partial}{\partial \bs \theta}
\left[
\rho_{\tau} \left( \frac{y_{it} - \mu_{it} (s_{it})
}
{\hat \sigma} \right) \right] = \tbf 0,
\end{align}
For the latent drop-out model, the score function is weighted by means of the LDO class posterior probabilities, $\hat \zeta_{ig}$, leading to
\begin{equation}
\sum_{i = 1}^n \sum_{g = 1} ^ {G-1} \hat \zeta_{ig}
\frac{\partial}{\partial \bs \lambda} 
\left\{
\log 
\left[
\left( \frac{e^{\lambda _{0g} + \lambda _{1} T_i} }{1+ e^{ \lambda _{0g} + \lambda _{1} T_i}} 
\right) 
-
\left( \frac{ e^{ \lambda _{0g-1} + \lambda _{1} T_i } }{1+ e^{ \lambda _{0g-1} + \lambda _{1} T_i}} 
\right)
\right]
 \right\} = \tbf 0
\end{equation}
The E- and the M- steps are repeatedly alternated until the difference between the likelihood values for two successive iterations is lower than a fixed constant $\epsilon>0$, that is 
\begin{equation*}
\ell^{(r+1)} - \ell^{(r)} < \epsilon.
\label{relDifference}
\end{equation*}
The algorithm reaches convergence for a given number of hidden states, $m$, and latent drop-out classes, $G$, which we consider fixed and known. For a given combination $[m,G]$, several starting points are used to avoid local maxima. As a result, we have a set of possible solutions, and the final $[m,G]$-based estimates come from the model with the highest log-likelihood value obtained over the set of starting points considered.
As it typically happens in the linear quantile mixed model framework, standard errors for parameter estimates are derived by exploiting a non-parametric block bootstrap \citep[see eg][]{Buchinsky1995}. Bootstrap samples are obtained by sampling individuals and retaining the corresponding longitudinal sequence, to preserve the within individual dependence structure.

\section{Real data example: CD4 data}\label{sec_realData}
To explore the empirical behaviour of the model, we consider the CD4 cell count data discussed, among others, by \cite{ZegerDiggle1994}.
These data come from the Multicenter AIDS cohort study (MACS) conducted since 1984 with the aim at analysing HIV progression over time (see \citealp{Kaslow1987}).
It includes nearly 5000 gay and bisexual men from Baltimore, Pittsburgh, Chicago and Los Angeles.
One of the effects of HIV is the reduction of T-lymphocytes, referred to as CD4 cells, which play a vital role in immune function; the virus progression can be assessed by measuring the number of CD4 cells over time.
\\ \indent
We have considered $2376$ repeated measurements coming from $369$ men who were seronegative at the beginning of the study and seroconverted in the meanwhile. They have been observed from $3$ years before up to $6$ years after the seroconversion: each individual has been followed from a minimum of $1$ to a maximum of $12$ occasions. While the time occasions are not equally spaced, the distribution of the time elapsed between successive visits is concentrated around $0.50$ (that is half a year) and, therefore, we may consider occasions as if they were equally spaced. This greatly simplifies notation and estimation.
At each visit, a number of covariates has been measured together with the level of T-lymphocytes in the blood: years since seroconversion (negative values indicate that the current CD4 measurement has been taken before the seroconversion), age at seroconversion (centered around $30$), smoking (packs per day), recreational drug use (yes or no), number of sexual partners, depression symptoms as measured by the CESD scale (larger values indicate more severe symptoms). The analysis has been conducted on the log transformed CD4 counts, that is log(1+CD4 count).
\\ \indent
As it is often the case with longitudinal designs, some of the units in the sample leave the study before its ending, and present incomplete information. In table \ref{tab1_idPerVisits}, we report the number of individuals available at each visit; as it can be seen, only a small number of individuals presents complete data records.
\begin{small}
\begin{table}[!htb]
\centering
\caption{Number of individuals in the study at each time occasion.}
\begin{tabular}{cccccccccccccccccc}
\toprule
\toprule
Visit & 1 & 2 & 3 & 4 & 5 & 6 & 7 & 8 & 9 & 10 & 11 & 12 \\ 
\midrule
& 369 & 364 & 340 & 315 & 268 & 225 & 173 & 133 &  92 &  54 &  33 &  10 \\ 
\bottomrule
\bottomrule
\end{tabular}\label{tab1_idPerVisits}
\end{table}
\end{small}

\begin{figure}[!ht]
\centering
\caption{Response variable distribution at each time occasion.}
\includegraphics[scale=0.30]{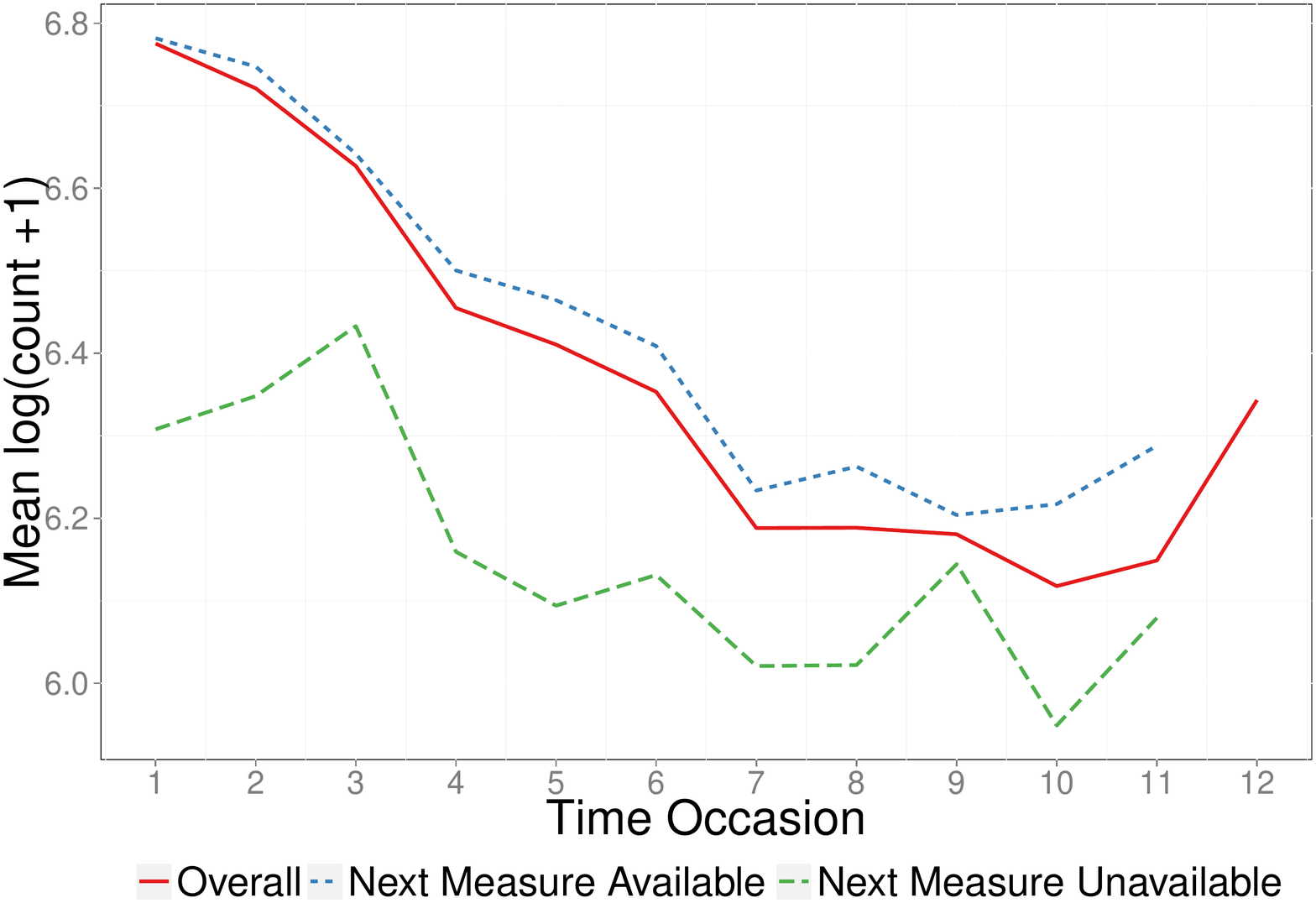} 

\label{mean_resp_overall_not}
\end{figure}

\begin{figure}[!htb]
\centering
\caption{LDO class probabilities for $\tau = 0.25$ (left) and $\tau = 0.50$ (right).}
\subfloat{
	\includegraphics[width = 0.44\textwidth]{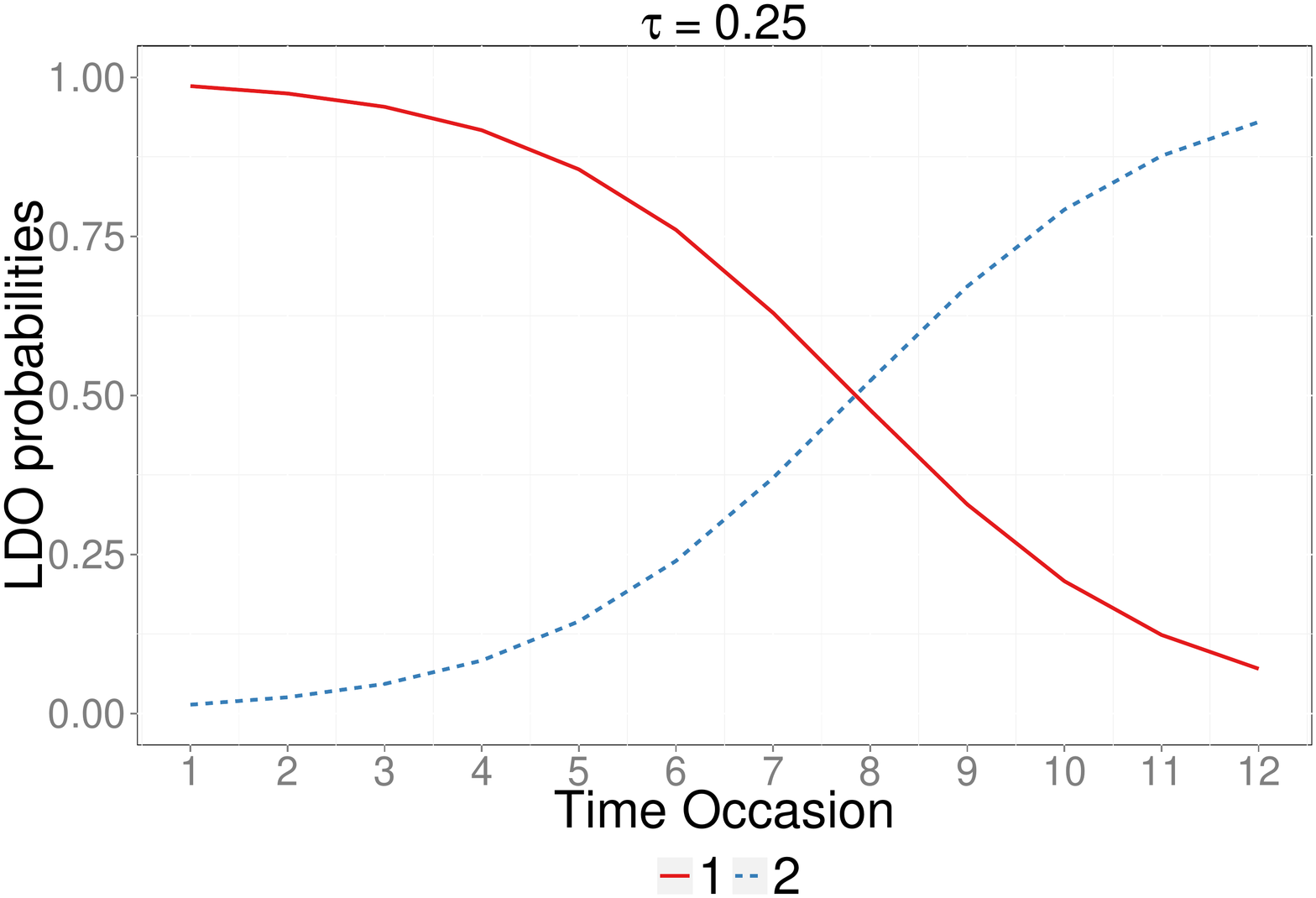}
	\label{q25_map}
}
\subfloat{
	\includegraphics[width = 0.44\textwidth]{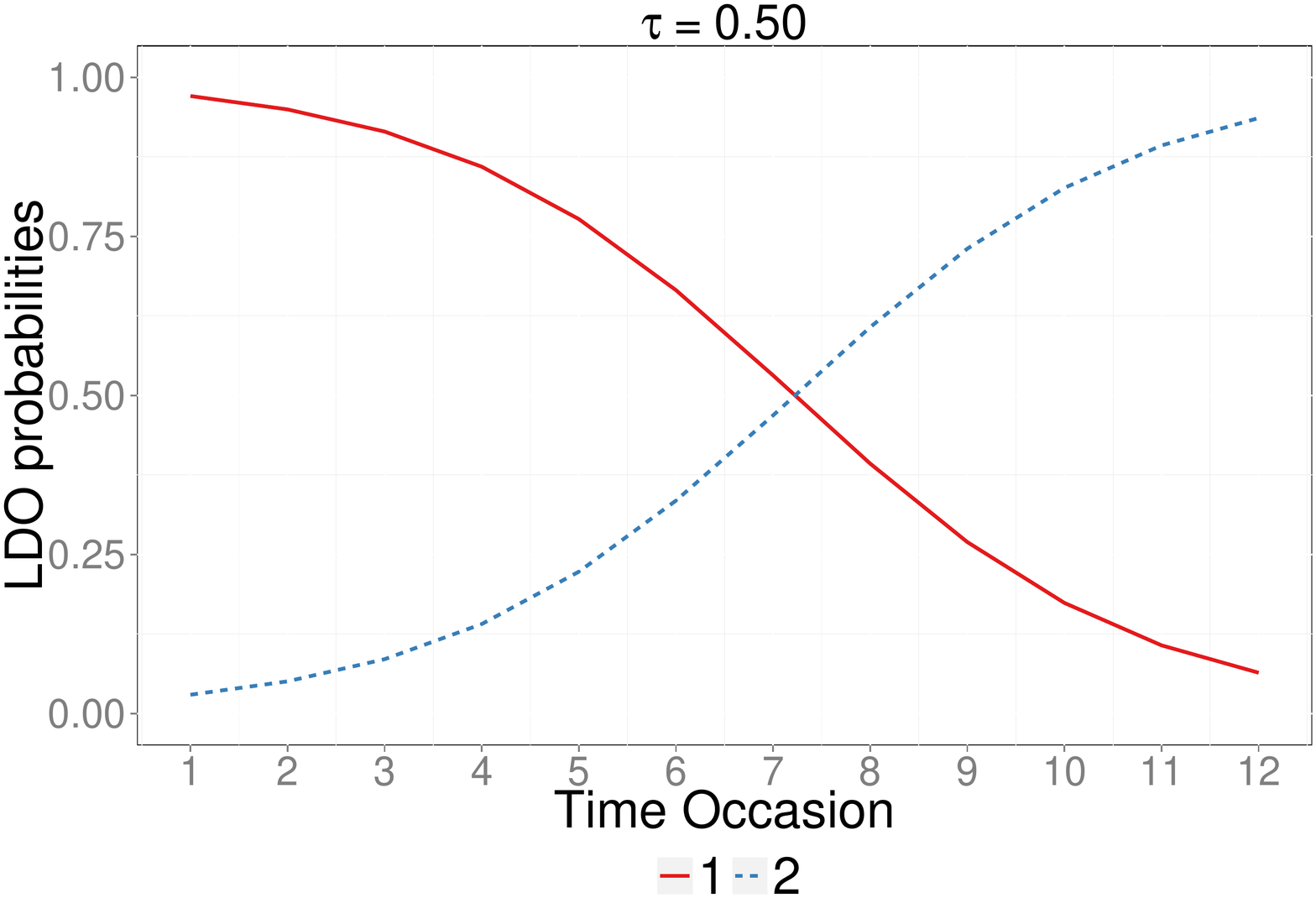}
	\label{q50_map}
}

\label{ldo_probs}
\end{figure}

Figure \ref{mean_resp_overall_not} displays the mean response evolution during the follow up, for the overall sample and stratified by whether or not the units drop-out from the study between the current and the subsequent time occasion.
As it may be noticed, a progressive decrease in the CD4 counts is observed, which is coherent with the progression of the virus.
However, some differences between the units staying and those dropping-out from the study between $t$ and $t+1$ may be noticed. The latter (individuals) present CD4 levels which are lower when compared to units remaining in the study beyond $t+1$, especially at the beginning of the observation window. 
These findings suggest the potential presence of some form of sample selection.
To analyse the effect of observed covariates on the HIV progression and account for the missing data process, we have estimated a linear quantile hidden Markov model with LDO-dependent transitions. To give some insight into the sensitivity of parameter estimates to modelling assumptions, we compare these results with those obtained from the corresponding MAR version, the lqHMM \citep{Farcomeni2012}. 
Being more severe HIV-related symptoms the main target of inference, we have decided to focus on lower CD4 count levels, that is on $\tau = (0.25, 0.50)$.
For a generic quantile $\tau \in (0,1)$, the following conditional model for longitudinal observations has been fit: 
\[
\mu_{it}(s_{it})= \alpha(s_{it}) + 
\tbf x_{it}^\prime \bs \beta 
\]
where $\alpha(s_{it})$ denotes a state-dependent random intercept, while $\tbf x_{it}$ includes two continuous covariates (years since seroconversion and age), the dummy variable drug (baseline: no) and three discrete variables (packs of cigarette per day, number of sexual partners and CESD score).
Both lqHMM and lqHMM+QLDO have been fit for a varying number of hidden states ($m = 2, ..., 5$) and, if the case, for a varying number of latent drop-out classes ($G = 2, ..., 5$).
To reduce the chance of being trapped in local maxima, we have adopted a multi-start strategy.  For the hidden Markov chain, a first deterministic starting solution has been obtained by setting prior and transition probabilities to $\delta_h = 1/m$ and $q_{kh} = (1 + s \mathbb I(h = k)) /(m + s),  h,k = 1, ..., m$ (for a suitable constant $s$) for all the LDO classes (when present). Parameters in the missing data model have been initialized by fitting an ordered logit to the response obtained by discretizing the distribution of the number of visits for each individual. To avoid singularities, a fraction $\xi$ of responses has been randomly perturbed. 
Initial values for the fixed longitudinal model parameters correspond to the maximum likelihood estimates of the linear quantile regression model for independent observations, while the time-varying random intercept has been initialized by adding Gaussian quadrature locations to the corresponding fixed intercept. Random starting values have been obtained by perturbing the deterministic ones. 
For each model (ie for each combination $[m,G]$), we have considered $30$ starting points and retained the solution with the highest likelihood. In table \ref{BIC_values}, we report the corresponding AIC and the BIC values for such solutions. 
\begin{table}[t]
\caption{Model selection; penalized likelihood criteria for different value of $m$ and $G$ at different quantiles.}
\centering
\begin{small}
\begin{tabular}{lrrrrrrrr}
\toprule
 \toprule
&\multicolumn{4}{c}{LDO classes}\\[0.12cm]
Hidden States  & \multicolumn{1}{c}{1} & \multicolumn{1}{c}{2} & \multicolumn{1}{c}{3} &\multicolumn{1}{c}{4} \\ 
\midrule
\multicolumn{8}{c}{$\mathtt{ \tau = 0.25}$}\\
 \hdashline[2pt/5pt]
\multicolumn{8}{c}{\textbf{AIC}}\\
 
 \hdashline[2pt/5pt]
2	& 3247.36	&	3215.99	&	3218.02	&	3231.74	\\
3	& 2895.26	&	2876.79	&	2870.91	&	2890.06	\\
4	& 2655.24	&	2642.60	&	2646.09	&	2656.89	\\
5	& 2550.21	&	2550.92	&	2556.75	&	2589.15	\\

 \hdashline[2pt/5pt]
\multicolumn{8}{c}{\textbf{BIC}}\\
 
 \hdashline[2pt/5pt]
2	& 3298.20	&	3278.56	&	3292.32	&	317.77	\\
3	& 2969.56	&	2978.47	&	2999.97	&	3046.49	\\
4	& 2760.83	&	2799.03	&	2853.36	&	2915.01	\\
5	& 2694.91	&	2777.74	&	2865.71	&	2980.23	\\

  \hline
\multicolumn{8}{c}{$\mathtt{ \tau = 0.50}$}\\
  \hdashline[2pt/5pt]
\multicolumn{8}{c}{\textbf{AIC}}\\
 
 \hdashline[2pt/5pt]
2	&2688.11	&	2664.24	&	2665.12	&	2672.56	\\
3	&2448.49	&	2432.94	&	2436.74	&	2450.87	\\
4	&2310.55	&	2305.78	&	2308.59	&	2337.79	\\
5	&2239.02	&	2242.75	&	2255.94	&	2282.33	\\

 \hdashline[2pt/5pt]
\multicolumn{8}{c}{\textbf{BIC}}\\
 \hdashline[2pt/5pt]
 
2	&2738.95	&	2726.81	&	2739.42	&	2758.59	\\
3	&2522.79	&	2534.62	&	2565.80	&	2607.30	\\
4	&2416.15	&	2462.22	&	2515.86	&	2595.90	\\
5	&2383.72	&	2469.57	&	2564.90	&	2673.41	\\

\bottomrule
\bottomrule
\end{tabular}

\end{small}
\label{BIC_values}

\end{table}
As it was expected, because of the high number of parameters in the lqHMM+QLDO formulation, both criteria suggest to retain the solution with $m = 5$ and $G=1$ for the quantiles we have considered. However, by looking at the AIC values, we noticed only slight differences between the solution $[m = 5, G =1]$ and $[m = 5, G =2]$. This suggests that, despite the highly parametrized structure of the lqHMM+QLDO formulation, model fit (as measured by the maximized log-likelihood value) is improved when accounting for the missing data generation process. Furthermore, simulation results in section \ref{sec_simul} show that the BIC leads, in most of the cases, to models with a lower (than the truth) number of LDO classes. Based on these findings, we will consider the model $[m = 5, G = 2]$ as the potential competitor for the MAR version (the lqHMM).
\\ \indent
Table \ref{longModel} reports the estimated parameters for the longitudinal data model under the lqHMM and the lqHMM+QLDO specifications, with corresponding $95\%$ confidence intervals (within brackets). The confidence intervals have been computed (using a block non-parametric bootstrap) with $B = 1000$ resamples.
\begin{table}[!ht]
\caption{Estimated parameters for the longitudinal data model at different quantiles.}
\centering 
\begin{footnotesize}
\scalebox{1}{
\begin{tabular}{lrrrrrrrrrr}
\toprule
\toprule
& \multicolumn{2}{c}{lqHMM} &&  \multicolumn{2}{c}{lqHMM+qLDO} \\
\midrule
  
 \multicolumn{6}{c}{$\mathtt{\tau = 0.25}$}\\
 \hdashline[2pt/5pt]
$\alpha_1$	&	4.738	&	(3.238	;	4.956)	&	4.728	&	(3.221		;	4.944)	\\
$\alpha_2$	&	5.699	&	(5.395	;	5.750)	&	5.693	&	(5.435		;	5.752)	\\
$\alpha_3$	&	6.126	&	(6.051	;	6.164)	&	6.118	&	(6.073		;	6.155)	\\
$\alpha_4$	&	6.509	&	(6.413	;	6.562)	&	6.500	&	(6.446		;	6.549)	\\
$\alpha_5$	&	6.843	&	(6.757	;	6.935)	&	6.832	&	(6.772		;	6.922)	\\

Age	&	0.001	&	(-0.006	;	0.005)	&	0.001	&	(-0.006		;	0.005)	\\
Drugs	&	-0.033	&	(-0.084	;	0.068)	&	-0.025	&(-0.074		;	0.062)	\\
Packs	&	0.082	&	(0.051	;	0.096)	&	0.082	&	(0.048		;	0.095)	\\
Partners	&	0.011	&	(0.002	;	0.018)	&	0.010	&	(0.000		;	0.017)	\\
CESD	&	-0.004	&	(-0.006	;	-0.001)	&	-0.004	&(-0.006		;	-0.001)	\\
$\text{Time}_{\text{sero}}$	&	-0.091	&	(-0.121	;	-0.075)	&	-0.089	&	(-0.121		;	-0.073)	\\

\hline
 \multicolumn{6}{c}{$\mathtt{\tau = 0.50}$}\\
  \hdashline[2pt/5pt]
$\alpha_1$	&	5.628	&	(5.074	;	5.753)	&	5.618	&	(5.142		;	5.751)	\\
$\alpha_2$	&	6.198	&	(6.014	;	6.252)	&	6.197	&	(6.060		;	6.233)	\\
$\alpha_3$	&	6.524	&	(6.393	;	6.574)	&	6.522	&	(6.450		;	6.558)	\\
$\alpha_4$	&	6.805	&	(6.719	;	6.874)	&	6.797	&	(6.753		;	6.854)	\\
$\alpha_5$	&	7.191	&	(7.084	;	7.291)	&	7.182	&	(7.112		;	7.271)	\\

Age	&	-0.003	&	(-0.007	;	0.005)	&	-0.003	&	(-0.007		;	0.005)	\\
Drugs	&	0.036	&	(-0.016	;	0.110)	&	0.038	&	(-0.007		;	0.082)	\\
Packs	&	0.049	&	(0.014	;	0.068)	&	0.048	&	(0.011		;	0.067)	\\
Partners	&	0.002	&	(-0.003	;	0.012)	&	0.001	&	(-0.004		;	0.011)	\\
CESD	&	-0.005	&	(-0.007	;	-0.001)	&	-0.005	&	(-0.007		;	-0.001)	\\
$\text{Time}_{\text{sero}}$	&	-0.110	&	(-0.126	;	-0.084)	&	-0.108	&	(-0.125		;	-0.080)	\\

\bottomrule
\bottomrule
\end{tabular}

}

\end{footnotesize}

\label{longModel}
\end{table}
As it can be easily noticed, age and drugs play no role in explaining the evolution of the CD4 cell counts over time. 
For both models, and for all the analysed quantiles, more severe depression symptoms lead to a decrease in the response variable; as expected, increases in the time since seroconversion correspond to a reduction in the level of T-lymphocytes. The effect of $\text{Time}_{\text{sero}}$ is slightly reduced under the lqHMM with respect to its MNAR counterpart. 
Results for the remaining covariates follow. Based on the results reported in table \ref{longModel}, smoking more cigarettes (for $\tau = 0.25$ and $\tau = 0.50$, with stronger effect in the former case) and having more sexual partners ($\tau = 0.25$ only) are associated to higher CD4 cell counts. According to \cite{ZegerDiggle1994}, the positive effect of such risk factors may be due to immune response stimulation or, simply, to a form of selection bias with healthier men staying longer in the study that continue their usual practices.
Regarding state-dependent intercepts, the estimates increase with the quantile level and, in all the analysed models, higher CD4 cell counts correspond to ``higher'' hidden states. 
When comparing results obtained under the lqHMM and the  lqHMM+QLDO, no substantial differences can be observed;  this suggests the class of models we are considering is rather robust with respect to possible misspecification of the missing data generating mechanism. However, when looking at the bootstrap confidence intervals, slight differences emerge. That is, if we consider the missing data process, we obtain narrower intervals and, therefore, improved reliability for parameter estimates.
By matching the results discussed so far with the estimated initial and transition probabilities, more thoughtful information on individual trajectories can be obtained. 
We report in table \ref{lqhmm} the parameters for the Markov chain  estimated under the lqHMM formulation.
For $\tau = 0.25$, it is clear that most of patients start the study with a medium/high level of CD4 cell counts ($\delta_3+\delta_4+\delta_5 >0.9$).
As the time passes by, the estimated $\textbf Q$ matrix highlights a high variability in the longitudinal trajectories. Transitions between states are quite likely; units being in lower hidden states  generally tend to move towards higher baseline values.
When analysing results we have obtained for the median response ($\tau = 0.50$), a different evolution of the response variable seems to be recovered.
Here, intermediate hidden states are the most likely at the beginning of the observation window ($\delta_2+\delta_3+\delta_4 >0.85$) and transitions between states are less frequent than that observed for $\tau = 0.25$ ($q_{hh}>0.8, \forall h = 1, ..., m)$. If any transition is observed, the probability of moving towards ``lower'' states is slightly higher than that of moving towards the highest ones.
\\ \indent
The analysis of results obtained under the lqHMM+QLDO specification can help understanding the effect of a potentially non-ignorable missingess on these results.
In figure \ref{ldo_probs}, we report the estimated LDO class probabilities obtained under the lqHMM+QLDO specification.
It may be noticed that, for both quantiles, higher classes are associated with increasing time to drop-out. That is, units staying longer into the study belongs to the second LDO class.
We report in tables \ref{lqhmm_ldo25}-\ref{lqhmm_ldo50} the estimates for the initial and the transition probabilities under the lqHMM+QLDO specification for the two classes (say $\text{LDO}_1$ and $\text{LDO}_2$).
Initial probability estimates, for all the analysed quantiles, suggest that the first hidden state is quite unlikely at the beginning of the study. Units are almost equally distributed over the remaining states.
As regards the transition probability matrices, parameter estimates highlight the presence of individuals in the sample who experience quite a different progression of the disease over time. Class $\text{LDO}_1$ is characterized by shorter individual sequences and mostly include subjects who leave the study prematurely. 
Within this class, the estimated transitions for $\tau = 0.25$ are quite similar to those observed for the lqHMM specification. Units with particularly low CD4 count levels move towards ``higher'' hidden states. The only remarkable difference between lqHMM and lqHMM+QLDO is related to $\hat q_{11}$ that, under the latter approach, is much higher ($\hat q_{11} = 0.931$ vs $\hat q_{11} = 0.798$). This is probably due to those units in the sample that leave the study with very low CD4 levels and that, under the MAR approach, are not clearly identified.
When we look at the results for $\tau = 0.50$, the estimated transitions suggest a progressive reduction in the median response over time. Comparing results obtained under the MNAR and the MAR approach, it is clear that such an evolution is better identified when accounting for the missing data process. In fact, under the LDO specification, the probability of moving towards the ``lowest'' state is higher than that observed for lqHMM and with probability equal to one individuals do not further move.
This result helps detect units that drop-out prematurely from the study after experiencing a steep and sudden reduction in CD4 count levels.
\\ \indent
Focusing on class $\text{LDO}_2$ (ie the class associated with units staying longer into the study) different longitudinal paths can be observed. When considering the left tail of the response distribution ($\tau = 0.25$), the first two hidden states are seldom visited and, if any transition is observed, units move towards ``higher'' states in at the next occasion. The only exception is for the estimate $\hat q_{31} = 0.184$ which is probably associated to some units that experience a sudden decrease in the CD4 level followed by an increase at the subsequent visit. Regarding the other hidden states, if any transition is observed, units generally tend to move towards higher baseline values.
A similar path can be observed for the median response, $\tau = 0.50$, where the estimated $\textbf Q$ matrix is almost diagonal, apart from the first hidden state which is, however, seldom reached. As for $\tau = 0.25$, also in this case, if any transition is observed, this is generally towards higher intercept values.
\\ \noindent
 To support the results we have discussed so far, 
we report in figure \ref{fig_ldo_map} the longitudinal trajectories of individuals classified (via a MAP criterion) into $\text{LDO}_1$ (left) and $\text{LDO}_2 $ (right), for $\tau = 0.25$ and $\tau = 0.50$.  
Local polynomial regression curves (blue lines), $95 \%$ confidence intervals (gray bands) and mean values (blue dots) are reported. Due to the missing data process, wider confidence intervals are observed at the last measurement occasions.
As expected, units in class $\text{LDO}_1$ leave the study earlier in time and experience a more evident reduction in the CD4 counts during the follow-up time. On the other hand, longer longitudinal sequences and more stable response patterns are observed for those units who are classified in $\text{LDO}_2$, for both $\tau = 0.25$ and $\tau = 0.50$.
\begin{figure}
\centering
\caption{Longitudinal trajectories by LDO class, for $\tau = 0.25$ (left) and $\tau = 0.50$ (right).}
\subfloat{
	\includegraphics[width = 0.49\textwidth]{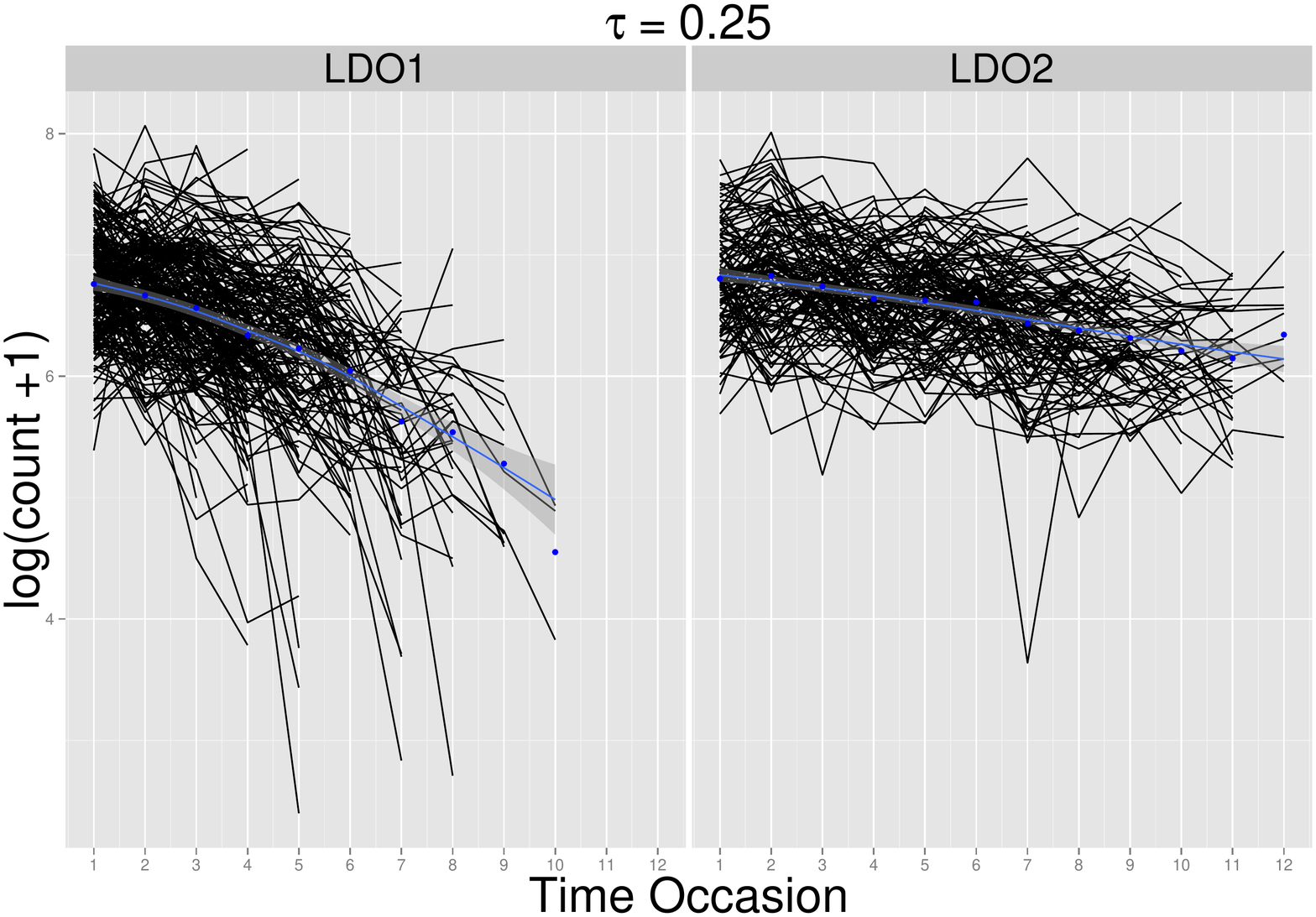}
	\label{fig_ldo_map}
}
\subfloat{
	\includegraphics[width = 0.49\textwidth]{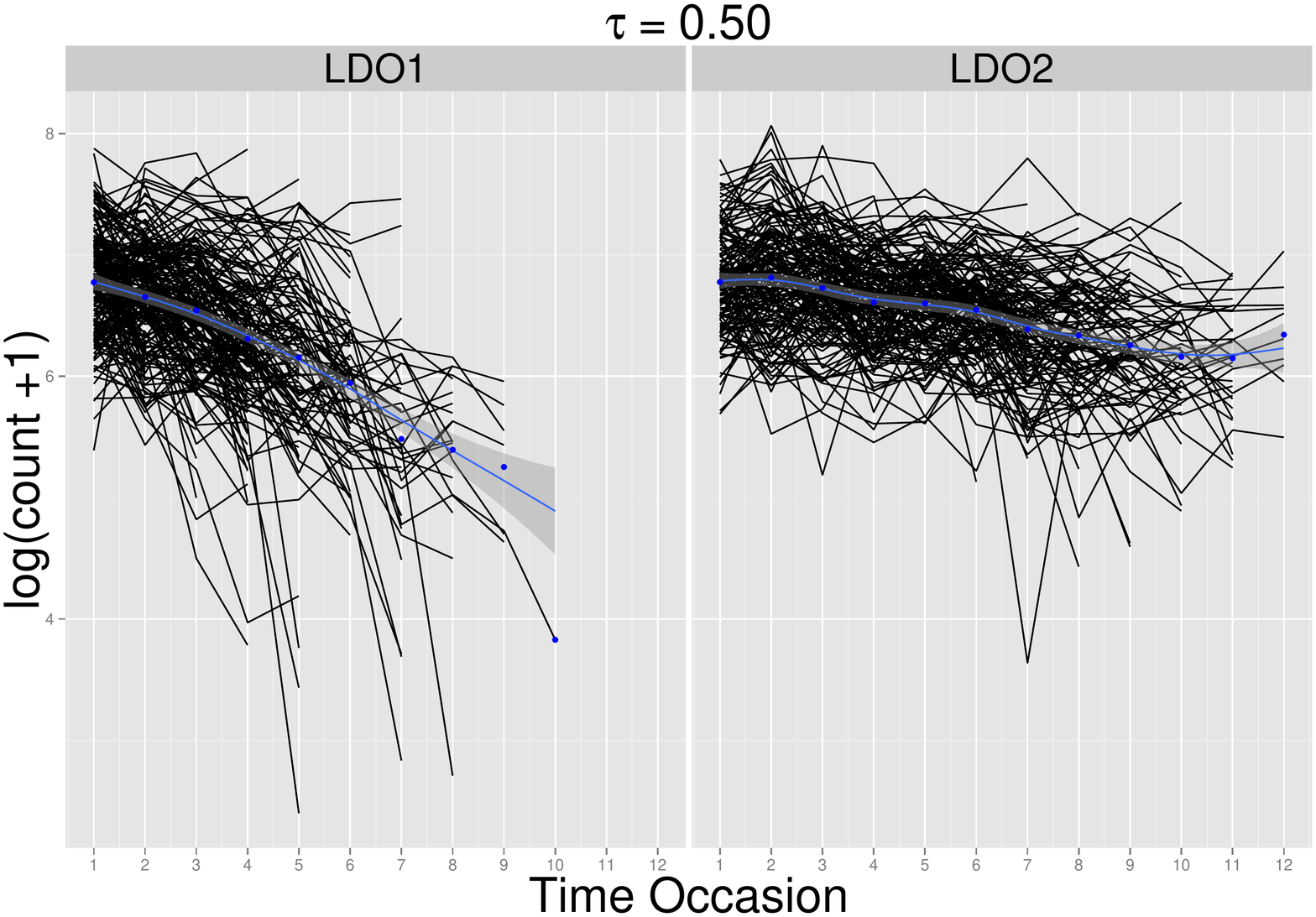}
	\label{q50_map}
}

\label{fig_ldo_map}
\end{figure}
While we can not postulate the proposed model is correct and the lqHMM is not (this is not our aim indeed), we may observe that, by considering an inhomogeneous hidden Markov representation due to a non random missing data generating process, some of the parameter estimates slightly change interpretation and we get a more complete and coherent picture of the response variable dynamics.

\section{Simulation study}\label{sec_simul}
To evaluate the empirical behaviour of the proposed model, we have performed the following simulation study. Data have been generated from a Gausian HMM+QLDO with $m = 4$ states and $G=2$ LDO classes. 
For the missing data model, we have considered the following set of model parameters: $\bs \lambda = (4.41, -0.63)$. Based on such values, ``higher'' LDO classes are associated to longer longitudinal sequences. Initial probabilities for the hidden Markov chain have been fixed to $\bs \delta = (0.05, 0.39, 0.48, 0.08)$, while transition probabilities have been set equal to 

\begin{center}
\begin{minipage}{.45\textwidth}
\raggedright
\[
\tbf Q(1) = \left(
\begin{array}{cccc}
1.00 &  0.00 & 0.00 &0.00 \\
0.27 & 0.73 &  0.00&0.00\\
0.00 & 0.23 & 0.71& 0.06\\
0.05 & 0.06& 0.00 & 0.89\\
\end{array}\right)
\]
\end{minipage}
\begin{minipage}{.45\textwidth}
\raggedleft
\[
\tbf Q(2) = \left(
\begin{array}{cccc}
0.91 & 0.09 & 0.00 & 0.00\\
0.05& 0.92 &0.03 &0.00\\
0.02 &0.03 & 0.94 & 0.01\\
0.00 & 0.00 & 0.01 & 0.99\\
\end{array}\right)
\]
\end{minipage}

\end{center}
Based on these parameter values, individuals belonging to the first LDO class move towards ``lower'' hidden states with a higher probability than units belonging to the second class. Here, we have decided to reduce the distance between the transition probability matrices associated to the LDO classes when compared to those estimated for the real data application. This has been done to verify the ability of the estimation algorithm in recovering the ``true'' latent structure.
As regards the longitudinal observations, covariates available for the CD4 dataset have been directly considered. The following set of fixed parameters has been considered: $\beta_\text{timeSero} = -0.088, \beta_\text{age} = 0.006, \beta_\text{drugs} = 0.148, \beta_\text{packs} = 0.055, \beta_\text{partners} = 0.009, \beta_\text{cesd} = -0.004$; 
on the other hand, state-specific random intercepts have been set to $\alpha = \{5.861, 6.306, 6.650, 7.039\}$. 
\\ \indent
Based on these parameters, we have simulated the response variable from a Gaussian distribution, with variance $\sigma^2 = 0.23$, corresponding to the variance for the ALD density estimated in the real data application for $\tau = 0.50$. Mean values have been defined according to the following model
\[
\mu_{it}(s_{it})= \alpha(s_{it}) + 
\tbf x_{it}^\prime \bs \beta,
\]
We have considered $B = 200$ samples and estimated a lqHMM+QLDO for different quantiles, $\tau = \{0.25, 0.50\}$, and for different choices of $m$ and $G$, $m = \{3,4,5 \}$ and $G = \{1, 2, 3\}$.
\\ \indent
The bias and the standard deviation of parameter estimates for the longitudinal data model estimated for fixed $m = 4$ and $G = 2$ are reported in table \ref{tab_simul}.
\begin{table}[ht]
\begin{center}
\caption{Bias and standard deviation of longitudinal model parameters for the lqHMM+QLDO with $m=4$ and $G=2$. $\tau = \{0.25, 0.50\}$}
\begin{tabular}{lrrrrrrrr}
\toprule
\toprule
&\multicolumn{2}{c}{$\tau = 0.25$}&& \multicolumn{2}{c}{$\tau = 0.50$}\\\cline{2-3}\cline{5-6}
& \multicolumn{1}{c}{Bias} &  \multicolumn{1}{c}{Sd} &&  \multicolumn{1}{c}{Bias} &  \multicolumn{1}{c}{Sd}  \\
\midrule
$\alpha_1$	&	0.0099	&	0.0027	&	&	0.0102	&	0.0013	\\
$\alpha_2$	&	0.0150	&	0.0008	&	&	0.0134	&	0.0034	\\
$\alpha_3$	&	0.0222	&	0.0018	&	&	0.0109	&	0.0018	\\
$\alpha_4$	&	0.0230	&	0.0204	&	&	0.0050	&	0.0075	\\
$\beta_\text{timeSero}$	&	-0.0017	&	0.0026	&	&	0.0011	&	0.0009	\\
$\beta_\text{age}$	&	-0.0004	&	0.0004	&	&	-0.0002	&	0.0001	\\
$\beta_\text{drugs}$	&	-0.0073	&	0.0027	&	&	-0.0118	&	0.0031	\\
$\beta_\text{packs} $	&	0.0005	&	0.0010	&	&	0.0002	&	0.0012	\\
$\beta_\text{partners}$	&	-0.0006	&	0.0007	&	&	0.0001	&	0.0003	\\
$ \beta_\text{cesd}$	&	0.0000	&	0.0001	&	&	0.0000	&	0.0001	\\
\bottomrule
\bottomrule
\end{tabular}\label{tab_simul}

\end{center}\end{table}
As it is expected, a higher bias is observed for the parameters related to the hidden Markov chain when compared to the fixed effect estimates. The quality of results reduces (that is bias and sd tend to increase) when considering the left tail of the response distribution as this represents a low density region with reduced information.
\\ \indent
We report in tables \ref{tab_simul_Q25}-\ref{tab_simul_Q50} the bias and the standard deviation (within brackets) of the estimated transition probability matrices for the LDO classes considering $\tau = 0.25$ and $\tau = 0.50$, respectively. For both quantiles, parameters are estimated with good accuracy in term of bias and (relatively) low variability, whatever the LDO class and the hidden state. 
\begin{table}[ht]
\caption{Simulation results. Bias and standard deviation (within brackets) of transition probability matrices for the lqHMM+QLDO with $m=4$ and $G=2$. $\tau = 0.25$}
\centering

\begin{tabular}{x{0.7cm}rrrrrrrrrrrr}
\toprule
 \toprule
 & \multicolumn{1}{c}{1} & \multicolumn{1}{c}{2} & \multicolumn{1}{c}{3} & \multicolumn{1}{c}{4} \\ 

\midrule

\multicolumn{2}{l}{$\mathtt{LDO_\text 1}$}\\
1&-0.002	(0.00)	&	0.002	(0.00)	&	0.000	(0.00)	&	0.000	(0.00)	\\
2&
-0.041	(0.01)	&	0.041	(0.01)	&	0.000	(0.00)	&	0.000	(0.00)	\\
3&
0.000	(0.00)	&	-0.074	(0.03)	&	0.062	(0.06)	&	0.011	(0.03)	\\ 
4&
0.002	(0.02)	&	-0.032	(0.02)	&	0.000	(0.00)	&	0.030	(0.03)	\\

\multicolumn{2}{l}{$\mathtt{LDO_\text 2}$}\\
1&0.017	(0.00)	&	-0.017	(0.00)	&	0.000	(0.00)	&	0.000	(0.00)	\\
2& 0.012	(0.02)	&	-0.009	(0.02)	&	-0.003	(0.00)	&	0.000	(0.00)	\\
3& 0.007	(0.01)	&	-0.006	(0.01)	&	0.004	(0.01)	&	-0.005	(0.00)	\\
4& 0.005	(0.00)	&	0.024	(0.01)	&	-0.004	(0.00)	&	-0.025	(0.02)	\\

\bottomrule
\bottomrule
\end{tabular}\label{tab_simul_Q25}

\label{lqhmm}

\end{table}

\begin{table}[ht]
\caption{Bias and standard deviation (within brackets) of transition probability matrices for the lqHMM+QLDO with $m=4$ and $G=2$. $\tau = 0.50$}
\centering
\begin{tabular}{x{0.7cm}rrrrrrrrrrrr}
\toprule
 \toprule
 & \multicolumn{1}{c}{1} & \multicolumn{1}{c}{2} & \multicolumn{1}{c}{3} & \multicolumn{1}{c}{4} \\ 

\midrule

\multicolumn{2}{l}{$\mathtt{LDO_\text 1}$}\\
1& -0.003	(0.00)	&	0.003	(0.00)	&	0.000	(0.00)	&	0.000	(0.00)	\\
2 & -0.042	(0.02)	&	0.042	(0.02)	&	0.000	(0.00)	&	0.000	(0.00)	\\
3& 0.007	(0.00)	&	-0.054	(0.02)	&	0.032	(0.03)	&	0.014	(0.01)	\\
4& -0.004	(0.01)	&	-0.034	(0.01)	&	0.000	(0.00)	&	0.038	(0.03)	\\

\multicolumn{2}{l}{$\mathtt{LDO_\text 2}$}\\
1 &0.027	(0.01)	&	-0.027	(0.01)	&	 0.000	(0.00)	&	0.000	(0.00)	\\
2& 0.015	(0.01)	&	-0.008	(0.01)	&	-0.007	(0.00)	&	0.000	(0.00)	\\
3& 0.004	(0.00)	&	-0.002	(0.01)	&	-0.001	(0.01)	&	-0.001	(0.00)	\\
4& 0.007	(0.00)	&	0.026	(0.01)	&	-0.004	(0.00)	&	-0.029	(0.01)	\\

\bottomrule
\bottomrule
\end{tabular}\label{tab_simul_Q50}

\label{lqhmm}

\end{table}
Last, in table \ref{tab_modelChoice}, we show the distribution of the estimated number of hidden states and LDO classes, using the AIC and the BIC criteria. As it is clear, AIC outperforms BIC in recovering the true number of states and classes. In fact, BIC tends to heavily penalize highly parametrized models. In the present context, for both quantiles, the BIC index suggests to adopt a lqHMM, that is a lqHMM+QLDO with a single LDO class (ie $G = 1$). On the contrary, AIC seems to recover with high accuracy the real model structure and it should be considered as a better choice to estimate $m$ and $G$. Surprisingly, when comparing $\tau = 0.25$ and $\tau = 0.50$, slightly better results with respect to the choice of $[m, G]$ are obtained in the former case. AIC always  identifies the right model for $\tau = 0.25$, while some anomalies have been observed for $\tau = 0.50$, where, in $11\%$ of samples, a further hidden state is selected. This is probably due to a more extreme behaviour in terms of state-specific locations which can be seldom observed at $\tau = 0.25$.

\begin{table}
\caption{Values of $m$ and $G$ estimated with BIC and AIC. $\tau = \{0.25, 0.50\}$.}
\begin{center}
\begin{tabular}{lrrrrrrrrrrr}

\toprule
\toprule
& \multicolumn{3}{c}{BIC} & & \multicolumn{3}{c}{AIC} \\
\cline{2-4} \cline{6-8}
& $G = 1$ & $G = 2$ & $G = 3$ &&  $G = 1$ & $G = 2$ & $G = 3$ \\
\midrule
\multicolumn{11}{c}{$\mathtt{\tau = 0.25}$}\\
\hdashline
$m = 3$ & 0.00 & 0.00 & 0.00 && 0.00 & 0.00 & 0.00\\
$m = 4$ & 0.99 & 0.01 & 0.00 && 0.00 & 1.00 & 0.00\\
$m = 5$ & 0.00 & 0.00 & 0.00 && 0.00 & 0.00 & 0.00\\

\hdashline
\multicolumn{11}{c}{$\mathtt{\tau = 0.50}$}\\
\hdashline
$m = 3$ & 0.00 & 0.00 & 0.00 && 0.00 & 0.00 & 0.00\\
$m = 4$ & 0.97 & 0.03 & 0.00 && 0.00 & 0.89 & 0.00\\
$m = 5$ & 0.00 & 0.00 & 0.00 && 0.00 & 0.11 & 0.00\\

\bottomrule
\bottomrule
\end{tabular}\label{tab_modelChoice}
\end{center}
\end{table}
To summarize, results we have obtained highlight the effectiveness of the estimation algorithm in recovering the ``true'', underlying, model structure. The quality of parameter estimates we have obtained in this simulation study suggests that the results presented in Section \ref{sec_realData} for the CD4 data analysis may be considered as quite reliable. The proposed model can be seen as a valid and flexible approach to handle informative  missing data patterns while controlling for time-varying sources of unobserved heterogeneity in longitudinal profiles. While the choice of letting $\tbf Q$ vary with the LDO class may lead to a substantial increase in the number of parameters, it may help describe the changes in the behaviour of units with a (possibly) different propensity to drop-out from the study.

\section{Conclusions}
Quantile regression models represent an interesting alternative to standard mean regression when the researcher's interest is on the tails of the response variable distribution and/or potential outliers in the data may affect the mean values. When responses are repeatedly measured over time on the same sample units, dependence between observations has to be taken into consideration to ensure the validity of inferential conclusions. 
In the presence of a potentially informative  missing data mechanism, standard statistical tools may lead to biased parameter estimates due to the ``selection'' of units remaining under observation.
In this paper, we have proposed a linear quantile hidden Markov model with drop-out dependent transitions. 
Within this framework, we obtain a more detailed picture of the response variable distribution and, jointly, address the problem of potentially non-ignorable missingness. 
More in detail, the latent drop-out class variable allows to capture (time-invariant) unobserved sources of heterogeneity shared by individuals with a similar propensity to drop-out. Such propensities lead to different transitions across the states of the hidden Markov chain; the marginal model for the longitudinal response is, therefore, given by a finite mixture of lqHMMs. 

We have re-analysed a benchmark dataset and compared the results obtained under the standard lqHMM by \cite{Farcomeni2012} with those from the proposed approach. Although with the proposed approach the number of parameters consistently increases, a clearer description of the observed data is obtained; this renders the proposed methodology an interesting and valuable alternative to existing modelling approaches.

\begin{landscape}
\begin{center}
\begin{table}[ht]
\caption{Estimated initial and transition probabilities at different quantiles for the lqHMM, $m = 5$.}
\centering
\begin{footnotesize}

\begin{tabular}{x{0.7cm}rrrrrrrrrrrr}
\toprule
 \toprule
 & \multicolumn{2}{c}{1} & \multicolumn{2}{c}{2} & \multicolumn{2}{c}{3} & \multicolumn{2}{c}{4} &  \multicolumn{2}{c}{5} \\ 

\midrule

\multicolumn{2}{l}{$\mathtt{\tau = 0.25}$}\\

$\delta$	&	0.002	&	(0.000	;	0.009)	&	0.033	&	(0.000	;	0.070)	&	0.333	&	(0.231	;	0.431)	&	0.426	&	(0.342	;	0.529)	&	0.206	&	(0.100	;	0.300)	\\[0.2cm]
1	&	0.798	&	(0.374	;	1.000)	&	0.040	&	(0.000	;	0.273)	&	0.129	&	(0.000	;	0.501)	&	0.000	&	(0.000	;	0.464)	&	0.033	&	(0.000	;	0.184)	\\
2	&	0.137	&	(0.067	;	0.208)	&	0.660	&	(0.436	;	0.778)	&	0.203	&	(0.090	;	0.429)	&	0.000	&	(0.000	;	0.029)	&	0.000	&	(0.000	;	0.020)	\\
3	&	0.004	&	(0.000	;	0.028)	&	0.137	&	(0.080	;	0.195)	&	0.689	&	(0.568	;	0.787)	&	0.155	&	(0.093	;	0.250)	&	0.015	&	(0.000	;	0.046)	\\
4	&	0.009	&	(0.000	;	0.017)	&	0.035	&	(0.000	;	0.070)	&	0.158	&	(0.100	;	0.232)	&	0.744	&	(0.656	;	0.808)	&	0.055	&	(0.021	;	0.109)	\\
5	&	0.000	&	(0.000	;	0.005)	&	0.008	&	(0.000	;	0.026)	&	0.045	&	(0.002	;	0.087)	&	0.050	&	(0.000	;	0.109)	&	0.896	&	(0.839	;	0.955)	\\

\multicolumn{2}{l}{$\mathtt{\tau = 0.50}$}\\

$\delta$	&	0.000	&	(0.000	;	0.000)	&	0.219	&	(0.060	;	0.310)	&	0.360&	(0.238	;	0.499)	&	0.326	&	(0.202	;	0.441)	&	0.095	&	(0.042	;	0.149)	\\[0.2cm]
																														
1	&	0.933	&	(0.802	;	1.000)	&	0.067	&	(0.000	;	0.198)	&	0.000	&	(0.000	;	0.000)	&	0.000	&	(0.000	;	0.000)	&	0.000	&	(0.000	;	0.000)	\\
2	&	0.068	&	(0.031	;	0.126)	&	0.847	&	(0.742	;	0.920)	&	0.085	&	(0.004	;	0.179)	&	0.000	&	(0.000	;	0.000)	&	0.000	&	(0.000	;	0.000)	\\
3	&	0.026	&	(0.000	;	0.066)	&	0.086	&	(0.030	;	0.163)	&	0.827	&	(0.718	;	0.902)	&	0.061	&	(0.002	;	0.135)	&	0.000	&	(0.000	;	0.002)	\\
4	&	0.002	&	(0.000	;	0.018)	&	0.065	&	(0.011	;	0.106)	&	0.032	&	(0.000	;	0.105)	&	0.861	&	(0.805	;	0.910)	&	0.040	&	(0.012	;	0.072)	\\
5	&	0.003	&	(0.000	;	0.017)	&	0.027	&	(0.000	;	0.070)	&	0.000	&	(0.000	;	0.045)	&	0.043	&	(0.000	;	0.115)	&	0.927	&	(0.857	;	0.983)	\\

\bottomrule
\bottomrule
\end{tabular}

\end{footnotesize}
\label{lqhmm}

\end{table}

\begin{table}[!ht]
\caption{Estimated initial and transition probabilities for the lqHMM+QLDO, $m = 5$, $G = 2$ and $\text{LDO}_1$.}
\centering

\begin{footnotesize}

\begin{tabular}{x{0.7cm}rrrrrrrrrrrr}
\toprule
 \toprule
 & \multicolumn{2}{c}{1} & \multicolumn{2}{c}{2} & \multicolumn{2}{c}{3} & \multicolumn{2}{c}{4} & \multicolumn{2}{c}{5}  \\ 

\midrule

\multicolumn{2}{l}{$\tau = 0.25$}\\

$\delta$	&	0.006	&	(0.000	;	0.019)	&	0.197	&	(0.153	;	0.232)	&	0.259	&	(0.228	;	0.292)	&	0.269	&	(0.243	;	0.300)	&	0.269	&	(0.224	;	0.322)	\\

1	&	0.931	&	(0.715	;	1.000)	&	0.069	&	(0.000	;	0.241)	&	0.000	&	(0.000	;	0.000)	&	0.000	&	(0.000	;	0.000)	&	0.000	&	(0.000	;	0.102)	\\
2	&	0.088	&	(0.049	;	0.132)	&	0.663	&	(0.525	;	0.767)	&	0.239	&	(0.124	;	0.384)	&	0.011	&	(0.000	;	0.053)	&	0.000	&	(0.000	;	0.000)	\\
3	&	0.015	&	(0.000	;	0.039)	&	0.144	&	(0.089	;	0.229)	&	0.704	&	(0.546	;	0.782)	&	0.137	&	(0.072	;	0.260)	&	0.000	&	(0.000	;	0.016)	\\
4	&	0.000	&	(0.000	;	0.020)	&	0.080	&	(0.011	;	0.153)	&	0.087	&	(0.011	;	0.250)	&	0.772	&	(0.576	;	0.860)	&	0.062	&	(0.001	;	0.151)	\\
5	&	0.005	&	(0.000		0.015)	&	0.021	&	(0.000	;	0.089)	&	0.123	&	(0.004	;	0.213)	&	0.042	&	(0.000	;	0.159)	&	0.809	&	(0.681	;	0.907)	\\

\multicolumn{2}{l}{$\tau = 0.50$}\\
		$\delta$	&	0.000	&	(0.000	;	0.004)	&	0.200	&	(0.073	;	0.299)	&	0.332	&	(0.200	;	0.479)	&	0.363	&	(0.229	;	0.449)	&	0.104	&	(0.062	;	0.153)	\\
																1	&	1.000	&	(0.917	;	1.000)	&	0.000	&	(0.000	;	0.083)	&	0.000	&	(0.000	;	0.000)	&	0.000	&	(0.000	;	0.000)	&	0.000	&	(0.000	;	0.000)	\\
2	&	0.138	&	(0.056	;	0.222)	&	0.793	&	(0.661	;	0.898)	&	0.069	&	(0.000	;	0.215)	&	0.000	&	(0.000	;	0.000)	&	0.000	&	(0.000	;	0.018)	\\
3	&	0.036	&	(0.000	;	0.118)	&	0.240	&	(0.088	;	0.404)	&	0.724	&	(0.183	;	0.846)	&	0.000	&	(0.000	;	0.461)	&	0.000	&	(0.000	;	0.000)	\\
4	&	0.000	&	(0.000	;	0.022)	&	0.116	&	(0.001	;	0.275)	&	0.123	&	(0.000	;	0.302)	&	0.721	&	(0.551	;	0.826)	&	0.040	&	(0.000	;	0.114)	\\
5	&	0.010	&	(0.000	;	0.039)	&	0.107	&	(0.000	;	0.223)	&	0.057	&	(0.000	;	0.277)	&	0.000	&	(0.000	;	0.364)	&	0.826	&	(0.527	;	0.955)	\\

\bottomrule
\bottomrule
\end{tabular}

\end{footnotesize}
\label{lqhmm_ldo25}

\end{table}

\end{center}

\begin{table}[!ht]
\caption{Estimated initial and transition probabilities for the lqHMM+QLDO, $m = 5$, $G = 2$ and $\text{LDO}_2$.}
\centering

\begin{footnotesize}
\begin{tabular}{x{0.7cm}rrrrrrrrrrrr}
\toprule
 \toprule
 & \multicolumn{2}{c}{1} & \multicolumn{2}{c}{2} & \multicolumn{2}{c}{3} & \multicolumn{2}{c}{4}  \\ 

\midrule

\multicolumn{2}{l}{$\tau = 0.25$}\\

$\delta$	&	0.006	&	(0.000	;	0.019)	&	0.197	&	(0.153	;	0.232)	&	0.259	&	(0.228	;	0.292)	&	0.269	&	(0.243	;	0.300)	&	0.269	&	(0.224	;	0.322)	\\

1	&	0.000	&	(0.000	;	0.000)	&	0.000	&	(0.000	;	0.000)	&	1.000	&	(1.000	;	1.000)	&	0.000	&	(0.000	;	0.000)	&	0.000	&	(0.000	;	0.000)	\\
2	&	0.000	&	(0.000	;	0.000)	&	0.000	&	(0.000	;	0.160)	&	0.000	&	(0.000	;	0.047)	&	0.726	&	(0.000	;	1.000)	&	0.274	&	(0.000	;	1.000)	\\
3	&	0.184	&	(0.000	;	0.994)	&	0.000	&	(0.000	;	0.000)	&	0.816	&	(0.000	;	1.000)	&	0.000	&	(0.000	;	0.124)	&	0.000	&	(0.000	;	0.000)	\\
4	&	0.007	&	(0.000	;	0.052)	&	0.064	&	(0.000	;	0.177)	&	0.046	&	(0.000	;	0.197)	&	0.763	&	(0.035	;	0.906)	&	0.121	&	(0.003	;	0.754)	\\
5	&	0.000	&	(0.000		0.000)	&	0.006	&	(0.000	;	0.168)	&	0.000	&	(0.000	;	0.024)	&	0.005	&	(0.000	;	0.150)	&	0.989	&	(0.765	;	1.000)	\\

\multicolumn{2}{l}{$\tau = 0.50$}\\
$\delta$	&	0.000	&	(0.000	;	0.004)	&	0.200	&	(0.073	;	0.299)	&	0.332	&	(0.200	;	0.479)	&	0.363	&	(0.229	;	0.449)	&	0.104	&	(0.062	;	0.153)	\\

1	&	0.515	&	(0.000	;	1.000)	&	0.485	&	(0.079	;	1.000)	&	0.000	&	(0.000	;	0.000)	&	0.000	&	(0.000	;	0.000)	&	0.000	&	(0.000	;	0.000)	\\
2	&	0.000	&	(0.000	;	0.062)	&	0.919	&	(0.438	;	1.000)	&	0.081	&	(0.000	;	0.721)	&	0.000	&	(0.000	;	0.032)	&	0.000	&	(0.000	;	0.000)	\\
3	&	0.018	&	(0.000	;	0.058)	&	0.011	&	(0.000	;	0.111)	&	0.900	&	(0.763	;	0.975)	&	0.071	&	(0.003	;	0.248)	&	0.000	&	(0.000	;	0.000)	\\
4	&	0.000	&	(0.000	;	0.020)	&	0.020	&	(0.000	;	0.055)	&	0.021	&	(0.000	;	0.089)	&	0.919	&	(0.862	;	0.968)	&	0.040	&	(0.000	;	0.087)	\\
5	&	0.000	&	(0.000	;	0.021)	&	0.000	&	(0.000	;	0.028)	&	0.000	&	(0.000	;	0.000)	&	0.039	&	(0.000	;	0.130)	&	0.961	&	(0.889	;	1.000)	\\

\bottomrule
\bottomrule
\end{tabular}
\label{lqhmm_ldo50}

\end{footnotesize}

\end{table}

\end{landscape}

\end{document}